\def\astroph{1}                              

\ifnum\astroph=0                             
\documentclass[12pt,preprint]{aastex}        

\else                                        
\documentclass[apjl]{emulateapj}
\usepackage{apjfonts}
\fi

\usepackage{epsfig,graphicx,natbib}
\usepackage{amsmath}
\newcommand{\vsh}{v_{\mathrm{sh}}}
\newcommand{\gad}{\gamma_{\mathrm{ad}}}
\newcommand{\wpet}{\omega_{pe}\,t}
\defcitealias{PicCode}{PCCP}
\slugcomment{Submitted to ApJ Letters}

\begin{document}
\shorttitle{3D modeling of collisionless shocks}
\shortauthors{T Haugb\o{}lle}

\title{Three dimensional modeling of relativistic collisionless ion-electron shocks}
\author{Troels Haugb\o{}lle}
\affil{Niels Bohr Institute}
\email{haugboel@nbi.dk}
\begin{abstract}
Two dimensional modeling of collisionless shocks has been of tremendous importance in
understanding the physics of the non-linear evolution, momentum transfer and particle
acceleration, but current computer capacities have now reached a point where three
dimensional modeling is becoming feasible. We present the first three dimensional model of
a fully developed and relaxed relativistic ion-electron shock, and analyze and compare it to
similar 2D models. Quantitative and qualitative differences are found with respect to
the two-dimensional models. The shock jump conditions are of course different, because of
the extra degree of freedom, but in addition it is found that strong parallel electric fields develop
at the shock interface, the level of magnetic field energy is lower, and the non-thermal particle
distribution is shallower with a powerlaw index of $\sim$2.2.
\end{abstract}

\maketitle
In collisionless shocks the mean free path of individual particles is much larger than
than the characteristic scales in the shock structure, and effective collisions and pressure
support are not mediated in particle-particle interactions, but by collective forces.
Collisionless shocks are ubiquitous in the universe, and happen on a range of scales
from the bow-shock of the solar wind at the earth, over solar corona, supernova remnants
to highly relativistic shocks at gamma-ray bursts and active galactic nuclei.
Given that only collective forces act as an effective collision mechanism, instabilities
can easily arise in the plasma. It has been shown over the last ten years how a range of Weibel-like
filamentation instabilities operate at the shock interface
\citep{Kazimura:1998,Medvedev:1999,Bret:2004,Frederiksen:2004,Silva:2003,Nishikawa:2003,Keshet:2009},
which can lead to magnetic field generation and particle acceleration.
A breakthrough for the study and understanding of collisionless shocks has been the
application of {\em ab initio} particle-in-cell simulations, where the formation and evolution beyond
the linear phase of the instabilities of the shocks can be studied with almost no assumptions.
Until now, the largest scale computer experiments of both magnetized
\citep{Shimada:2004,Saito:2004,Hededal:2005a,Spitkovsky:2005,Sironi:2009} and
unmagnetized \citep{Haugboelle:2005,Hededal:2005b,Spitkovsky:2005,Silva:2006,Kato:2007,
Chang:2008,Spitkovsky:2008a,Spitkovsky:2008b,Martins:2009} collisionless shocks have mostly
been performed in two dimensions, while three dimensional simulations of unmagnetized shocks
have been too costly to scale to large enough sizes. Some of the first studies of electron-positron
and electron-ion shocks were performed in 3D, but only the very early linear and quasi-linear
stages of the shock formation could be followed \citep{Frederiksen:2004,Silva:2003,Nishikawa:2003}.
Three dimensional studies of pair plasmas have
been done showing the full development of the shock ramp, and recovering the correct jump
conditions \citep{Haugboelle:2005,Spitkovsky:2005,Nishikawa:2009}, but only
limited electron-ion simulations have been made \citep{Spitkovsky:2008a}.

The development of long term large scale two dimensional simulations have helped
tremendously in the understanding of collisionless shocks, both in quantitative and qualitative
terms, but given current computational resources, and the technical quality of the Particle-In-Cell codes,
we now have the possibility of using fully three dimensional modeling to properly account
for the full dynamics.
In this paper we present the first three dimensional simulation of a relativistic unmagnetized collisionless
electron-ion shock, where the simulation is followed long enough to establish the correct jump
conditions, create a fully thermalized and developed downstream region, and follow the
emergence of a powerlaw distributed population of high energy ions and electrons downstream of the shock.
We compare the results to similar 2D simulations, and show how the different dimensionality impacts on
the formation and evolution of the shock.

\begin{figure*}[!ht]
\begin{center}
\includegraphics[width=0.96\textwidth]{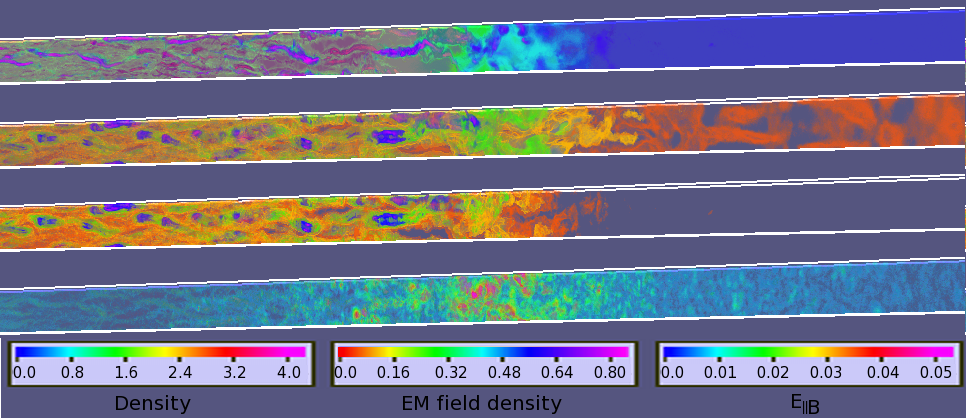}
\end{center}
\begin{center}
\includegraphics[width=0.96\textwidth]{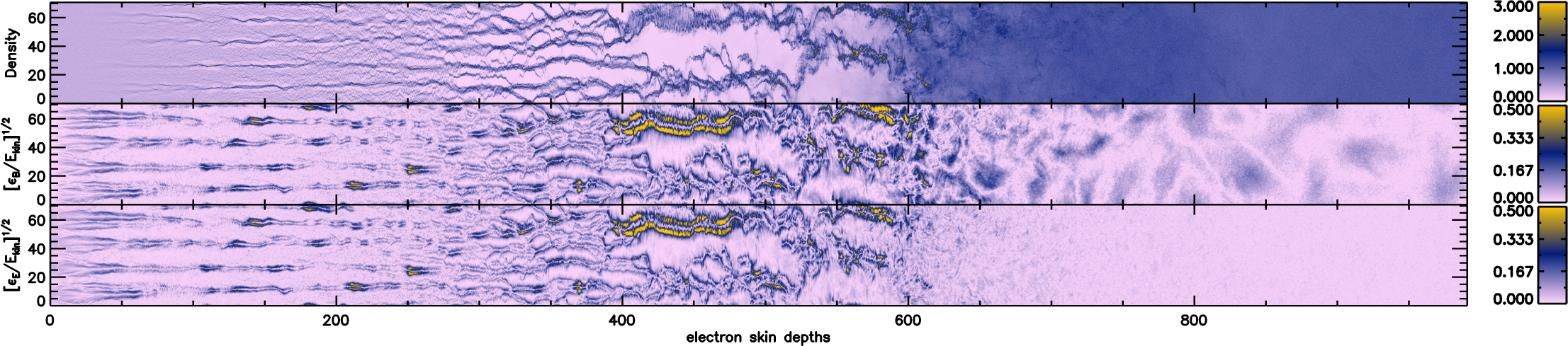}
\end{center}
\vspace*{-0.5cm}
\caption{Upper panel: From top to bottom the ion density, the magnetic field density $\epsilon_B^{1/2}$,
electric field density $\epsilon_E^{1/2}$, and the electric field projected along the magnetic field
lines $E_{\parallel B}$ at $\wpet\,=2250$. The fields are normalized to the kinetic energy
density of the upstream bulk flow, as for example $\epsilon_B = B^2 / [2\,n\,(m_i + m_e)\,(\Gamma-1)]$.
Lower panel: From top to bottom the ion density (normalized to 1/3 upstream), the magnetic field density,
and the electric field density at $\wpet\,=1500$ for a 2D collisionless shock.
}\label{fig:fields}
\end{figure*}
\section{Simulation setup}
We have used the massively parallel 3D particle-in-cell
\emph{Photon-Plasma} code, developed at the Niels Bohr Institute
\citep{Haugboelle:2005,Hededal:2005b}, to
simulate collisionless electron-ion shocks. For the three dimensional ion-electron run
we used a 2$^\mathrm{nd}$ order field solver and TSC interpolation of particles, while for
all other runs we used a newly developed 6$^\mathrm{th}$ order field solver and cubic
interpolation of particles, giving effectively a 50\% higher resolution
\citepalias[see e.g.~][]{PicCode}. The simulation domain
is periodic perpendicular to the streaming direction, while the lower inflow boundary is
open for outgoing particles and has absorbing field boundaries and the opposite, upper
boundary is a perfectly reflecting wall. We use a combination of a moving injector,
to launch the shock \citep{Sironi:2009}, and a moving window, to follow
the shock evolution \citep{movingframe}. Initially, only a small slice in front of the wall is populated
with a streaming plasma, and a shock is launched when the reflected fields and
particles collide with the inflowing plasma. New particles are added to the
plasma in front of the shock with the speed of light until the full box is
populated with a streaming plasma. At the lower boundary outgoing particles are
allowed to escape the box, and electromagnetic fields are damped while the shock transition
propagates through the box and the downstream region grows. When the shock
transition reaches the center of the box, a moving frame is applied which continuously
moves the simulation domain to keep the shock transition at the center.
Initially the plasma is unmagnetized, and hence nearly transparent for the particles.
The particles reflected by the wall creates an artificial counter-streaming population of particles
upstream of the shock transition region. While in 2D simulations the domain can be made
large enough for this initial spike of particles to diminish in density and become
insignificant, the required domain sizes are currently prohibitively expensive in 3D. 
But with the combination of the open lower boundary and the moving frame technique we can nevertheless follow
the evolution of the shock until it has settled to a steady state, with a thermalized population
downstream of the shock, and an in-streaming population mixed
with particles reflected from the shock upstream.
All runs have been done with an upstream Lorentz factor of $\Gamma=15$. The 3D simulation
was done with 250x250x7000 cells, 6 particles per species per cell,
a mass ratio $m_i/m_e$ of 16, and contained up to 16 billion particles. To properly
resolve the dynamics both up- and downstream of the shock we resolve the electron skin
depth $\delta_e = (4 \pi n_e e^2 / m_e \Gamma)^{1/2}$ with 14 grid cells upstream and
7 cells downstream of the shock, corresponding to 56 and 23 cells per ion skin depth up- and
downstream of the shock. For the time stepping we used a Courant condition of 0.4,
checking both light crossing time in a grid cell, and the local plasma frequency. To check
the impact the box size had on the evolution we ran 2D simulations with initially 
250x7000 cells and then wider and longer boxes with up to 1000x14000 cells.
Additionally, to probe a larger dynamic range in 3D we also performed a pair-plasma simulation with
375x375x3750 cells and 7.5 cells / skin depth, for a volume of 50x50x500 skin depths.

\section{Shock structure and evolution}
\begin{figure}[th]
\includegraphics[width=0.45\textwidth]{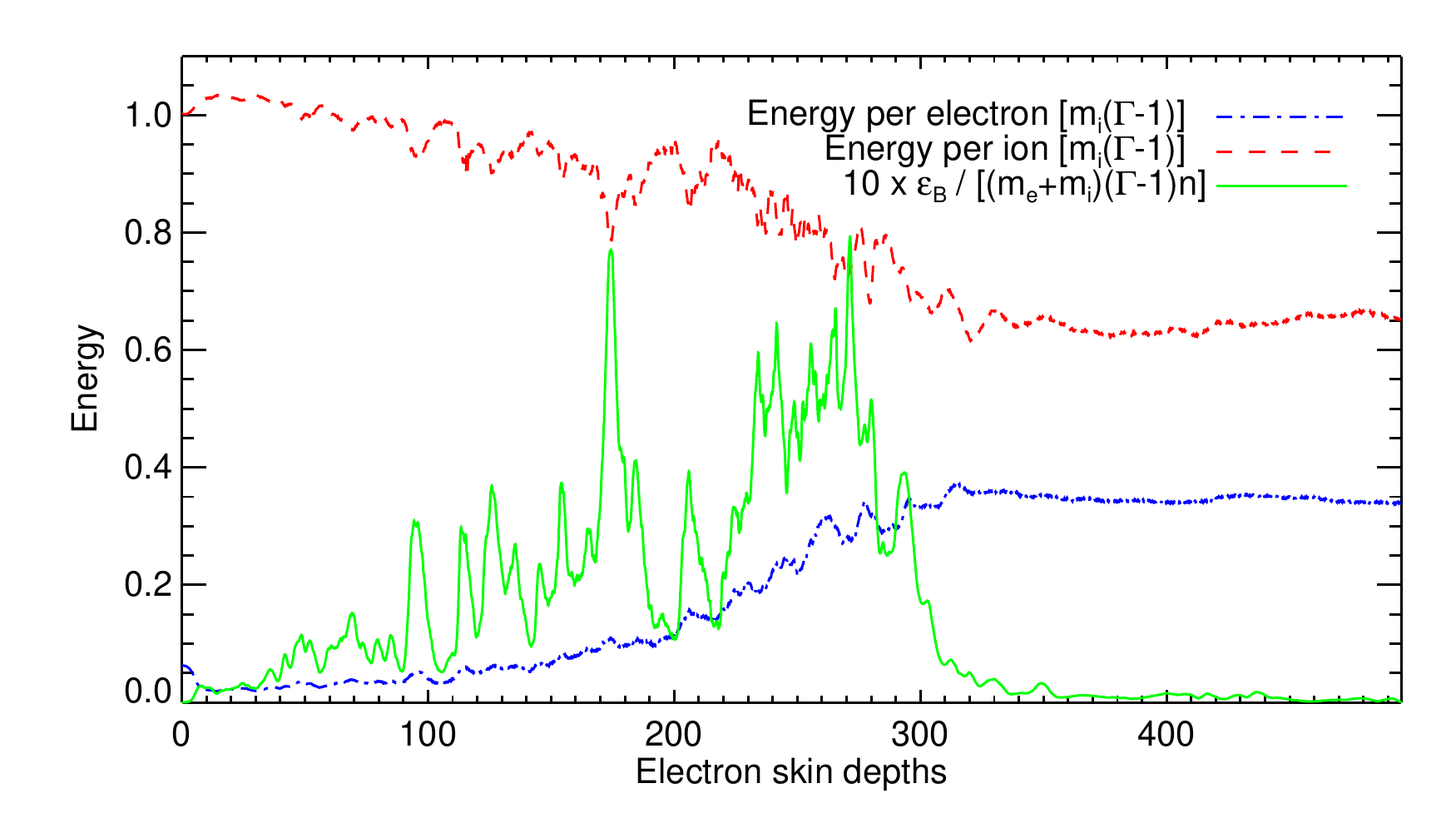}
\caption{Average energy distribution across the shock at $\wpet=2250$. To
avoid biasing only particles with positive velocities are used in calculating the average
kinetic energy per particle, and the magnetic field density is normalized to the in-streaming
bulk kinetic energy.}\label{fig:avenergy}
\end{figure}

Figure \ref{fig:fields} shows the structure of the evolved 3-D shock at $\wpet=2250$.
Upstream of the shock the particles reflected at the shock interfaces interact with the
incoming particles, generating a two-stream instability where the ions collect into
current channels and are pinched by the self-generated magnetic field. The electrons
Debye-shield the ions while oscillating in the strong transverse electric field, giving
effective electron heating and momentum transfer from ions to electrons. The current
channels slowly merge while loosing momentum, and break up at the shock interface.
The behavior of the upstream medium in the 3D simulation is in qualitative agreement
with the 2D simulations, though the current channels in 3D seem to merge less than in 2D. This
may be due to the fact that current channels in 2D (being 1D curves) necessarily cross, while in 3D
that is not the case (compare e.g. densities in fig. \ref{fig:fields}).
The average $\epsilon_B$ and the level of
electron to ion momentum are similar in 2D and 3D (see figure \ref{fig:avenergy}),
while the average $\epsilon_B$ with a maximum at 8\% is slightly lower in 3D.
\begin{figure}[th]
\includegraphics[width=0.45\textwidth]{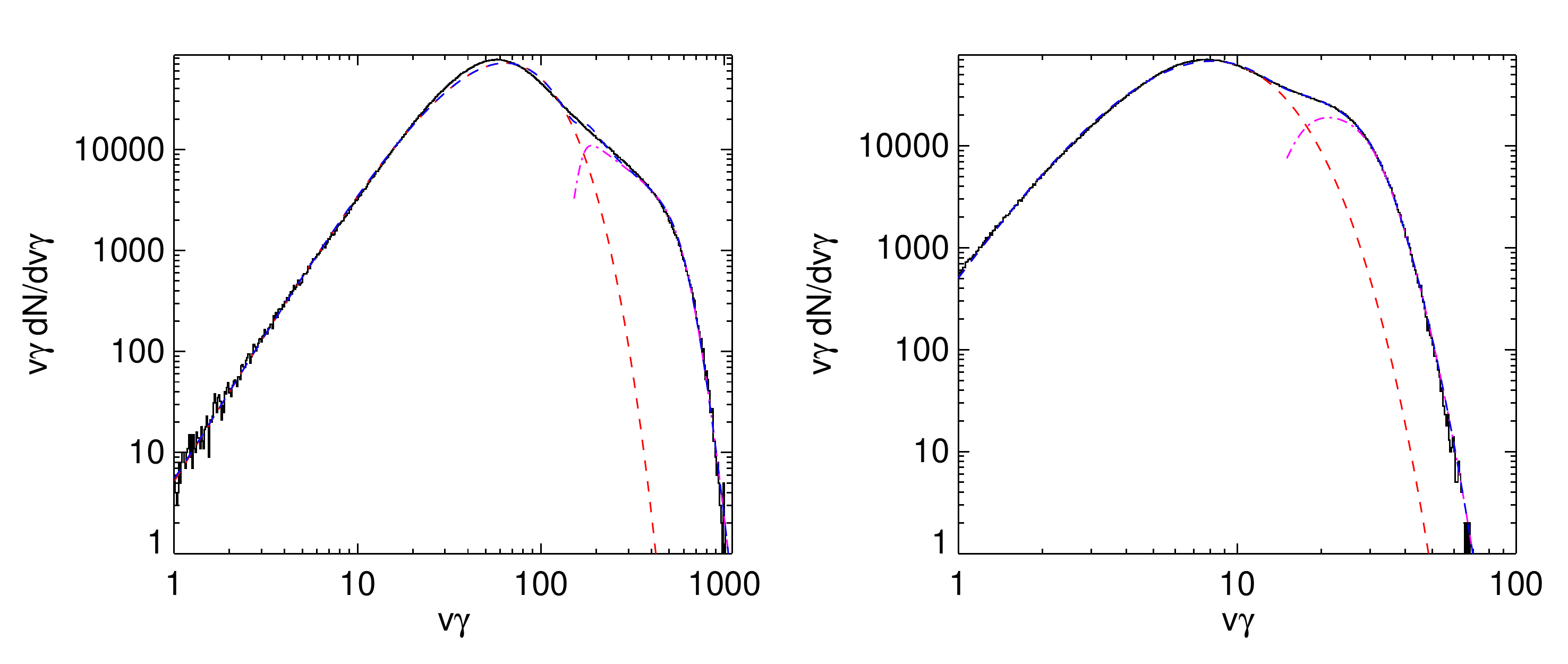}
\caption{Particle distribution function sampled downstream of the shock above the dashed
line marked on figure \ref{fig:phasespace} at $\wpet=2250$. To the left are electrons and to
the right ions. The red dashed line is a relativistic Maxwellian, the purple line is a powerlaw
with a low and high-$\gamma$ cut-off, and the blue line is the combined model.}\label{fig:pdf}
\end{figure}
Eventually the highly intermittent electromagnetic fields in the upstream current channels have
transferred sufficient bulk flow energy to heat and transverse momentum to make the channels 
diffuse and slow down. A transition to a downstream shocked medium occurs over approximately 20
ion skin depths, equivalent to about one Larmor radius (given that  the ions and electrons have approximately
the same energy, their Larmor radii are similar).
While the width of the shock transition is nearly the same in 2D and 3D, unique for 3D is
the occurrence of large 
electric fields parallel with the magnetic field (see upper panel in figure \ref{fig:fields})
in the transition region, coupled with $\epsilon_B=1$ locally,
giving an effective mechanism for ion and electron acceleration across the shock transition. A
similar phenomena has been observed with satellites in the collisionless shocks in the aurora
and the Earth's magnetosphere, where parallel electric fields together with strong particle acceleration
are observed at the shock transition \citep[see e.~g.\ ][]{Ergun:2001,Ergun:2009}. We also note that
parallel electric fields play a key role in reconnection of magnetic fields
\citep{Hesse:1988}, such as the reconfiguration that happens here, going from the mostly transverse current driven
axial magnetic field upstream of the shock to the turbulent flux ropes seen downstream of the shock.
In the downstream region the plasma is very nearly neutral, with little density variation, but with a high level of
magnetic turbulence. Our current simulation box for the ion-electron shock is not wide enough to allow
for the largest magnetic structures, but using a 3D pair plasma simulation with a larger transverse volume we
have observed how closed flux ropes are formed and are advected downwards from the shock, similar to
what was seen by \citet{Spitkovsky:2005}.

\section{Particle acceleration and distribution}
In 2D models of collisionless shocks it has
been found that particles are slowly accelerated by scattering off the filaments
\citep{Hededal:2004,Spitkovsky:2008b,Martins:2009}. A few ``lucky'' particles cross
back and forth over the shock interfaces a number of times, and this Fermi-like acceleration
process slowly builds up a power-law tail of high energy particles, due to the quasi constant
probability that a single high-energy particle is reflected in the strong transverse electric field
near the shock \citep{Martins:2009}.  Even though the acceleration mechanism is not
identical to the normal Fermi process, the resulting power-law index is in good agreement with
theoretical predictions, and is approximately $\alpha=$2.3 -- 2.6, where
$f(p) \propto \gamma^{-\alpha}$ at high energies. In our 3D simulation we find the same
emergence of a high energy tail distribution, on top of a relativistic Maxwellian downstream of
the shock. We model it as
\begin{eqnarray} \nonumber
f(v\gamma)\, = \, A_1 (v\gamma)^2 \exp\left(-m_{i,e}\gamma / T\right)
  + A_2 \gamma^{-\alpha} \\ \label{eq:pdf}
\qquad\qquad [1 + \exp(-(\gamma_{min} - \gamma)/\Delta_{min})]^{-1} \\ \nonumber
\qquad\qquad [1 + \exp(-(\gamma - \gamma_{max})/\Delta_{max})]^{-1}
\end{eqnarray}
where $A_i$ are normalizations, $T$ is the temperature, $\alpha$ is the powerlaw slope,
and $\gamma_i$ \& $\Delta_i$ are the locations and widths of the cut-offs. We impose the low
energy cut-off so the power-law makes a smooth match to the Maxwellian, while the upper
cut-off is time-dependent and grows with time. We find that in the 3D simulation the slope
is shallower with $\alpha=2.1-2.3$, matching the theoretical prediction of \citet{Keshet:2005},
for a $\Gamma=15$ relativistic shock of $\alpha_{theo}=2.1$.

\begin{figure}[!ht]
\includegraphics[width=0.45\textwidth]{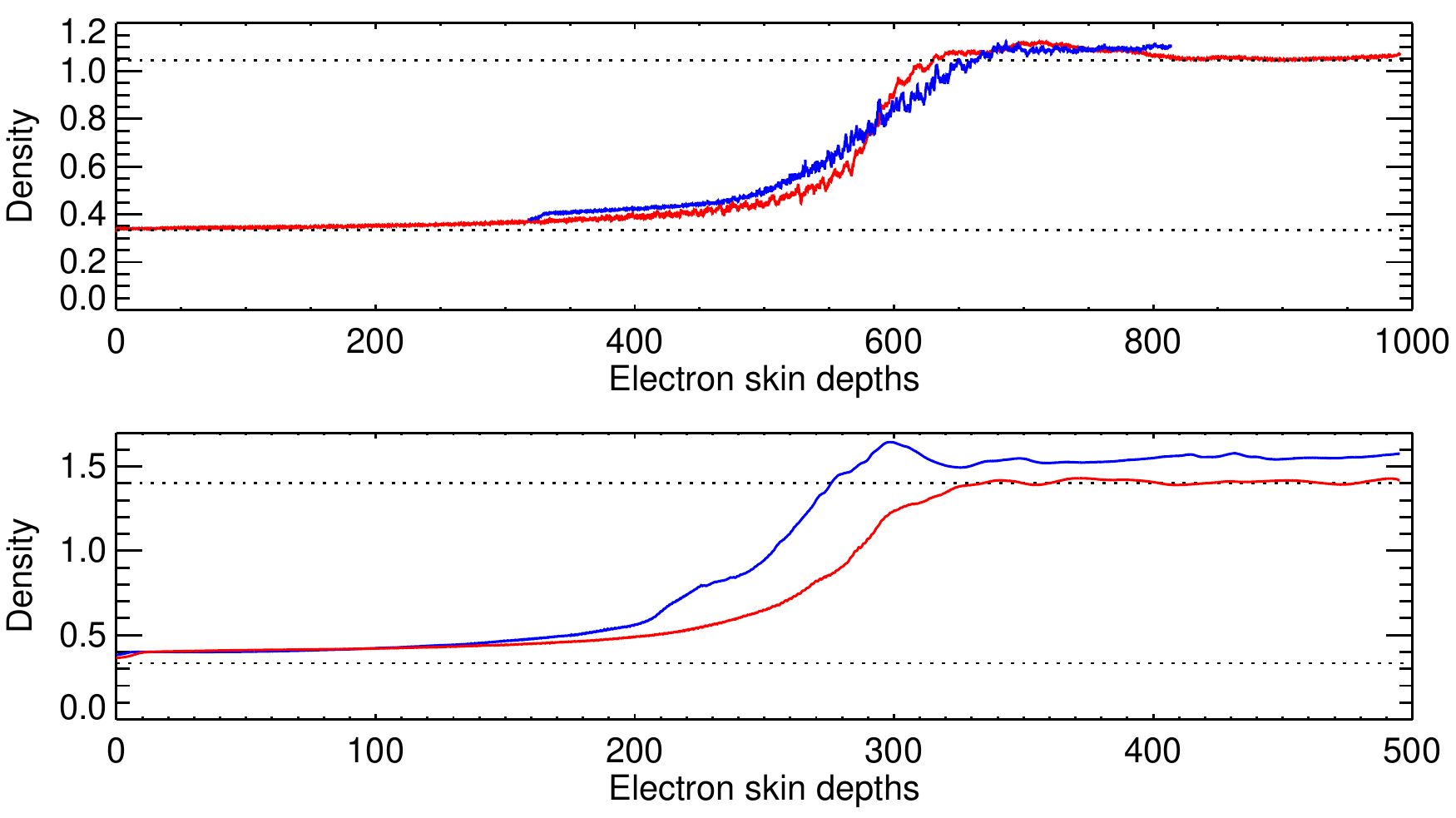}
\caption{Density profile of the 2D and 3D runs. In the upper panel the
density profile at $\wpet=1600$ is shown for a 14000x1000 cell domain,
and a 7000x250 cell domain. In the lower panel the density profile for the 3D run a the times
$\omega_{p} t=1500$ (red line) and $\wpet=2250$ (blue line) is shown.}\label{fig:dens}
\end{figure}

\section{Convergence}
The density and momentum distribution are some of the lowest order criteria to check when
modeling a collisionless shock. To assess the impact of our relatively limited simulation domain 
we compared the different 2D runs, finding that the limited box size has only a minor impact
on the jump conditions and on the evolution of the filaments. Analyzing the 2D runs we find
that a box with 250x7000 cells contains a shock with velocity $\vsh=0.42c$, and a downstream to
upstream density ratio of $n_d/n_u=3.24$, while a box with twice the length, independent of the
transverse size, more faithfully reproduces the analytic jump conditions.  With
$\vsh=0.47c$ and a density jump of 3.12 it is in percent precision agreement with the analytic expectation
for a relativistic gas $n_d/n_u = \gad / [\gad - 1] + 1 / [\Gamma (\gad - 1)]$, where $\gad$ is the
adiabatic index of the gas, which is 4/3 (3/2) for a 3D (2D) relativistic gas. This is also seen in
the 3D case, where we find $\vsh=0.27$ and $n_d/n_u = 4.62$, where analytically one
expects $\vsh=0.31$ and $n_d/n_u=4.2$ (see figure \ref{fig:dens}). Because we launch the shock
\begin{figure}[!th]
\includegraphics[width=0.45\textwidth]{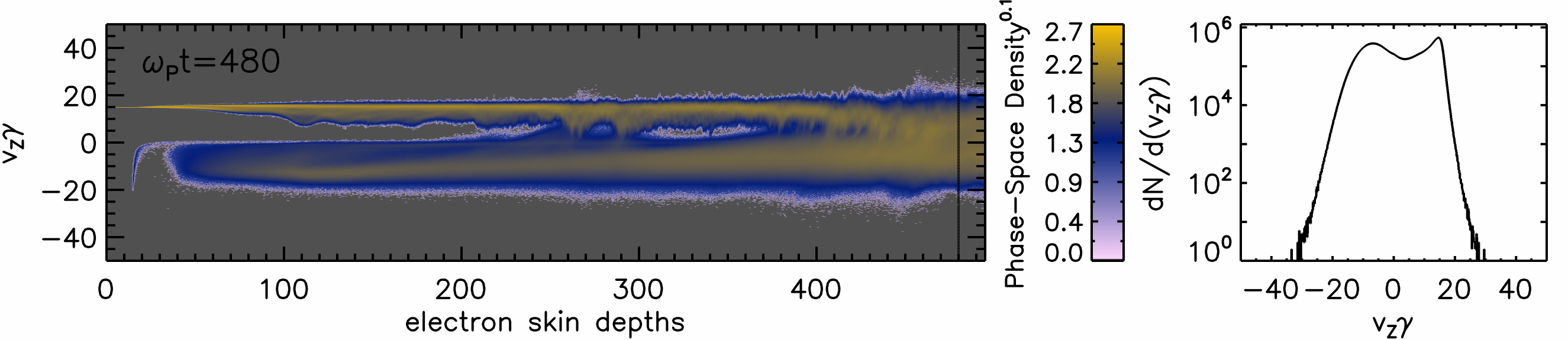}
\includegraphics[width=0.45\textwidth]{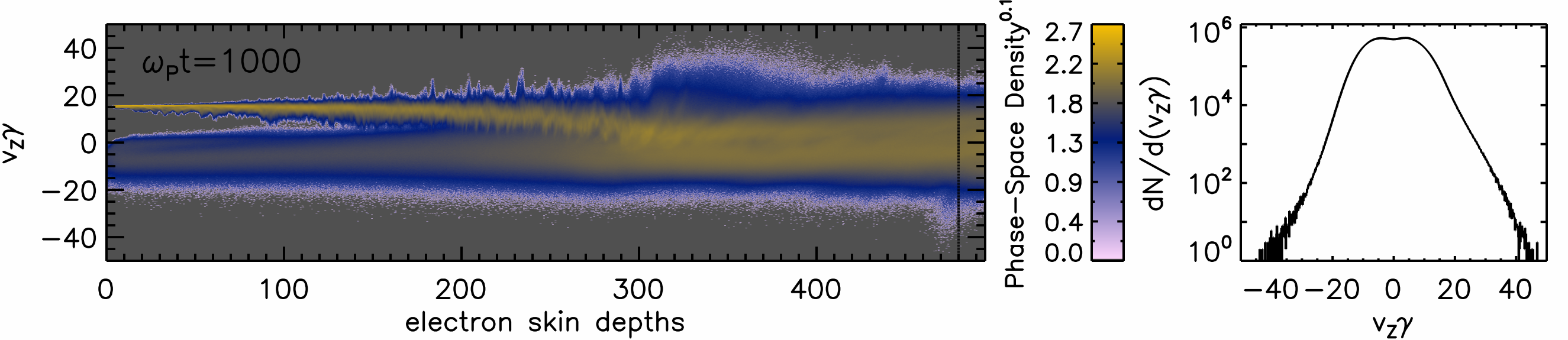}
\includegraphics[width=0.45\textwidth]{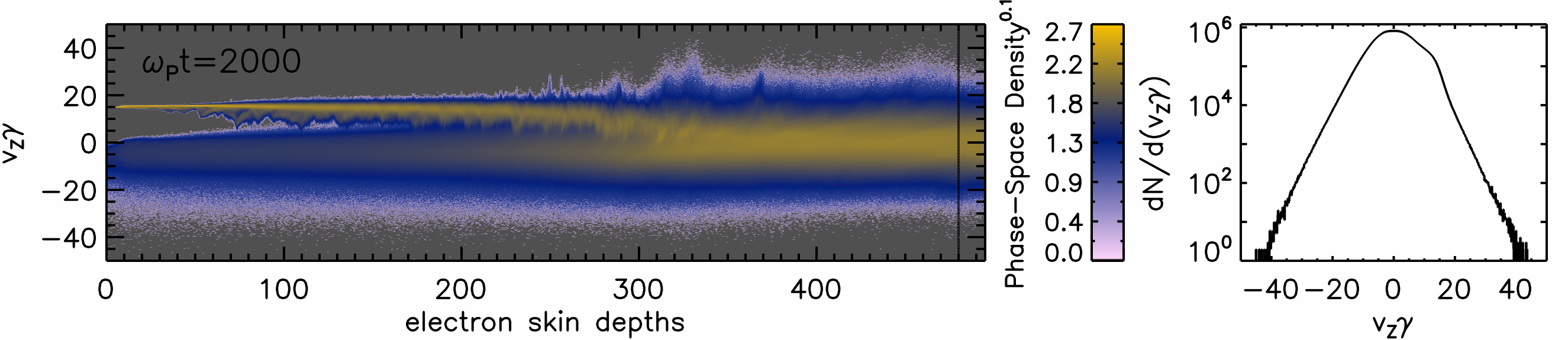}
\includegraphics[width=0.45\textwidth]{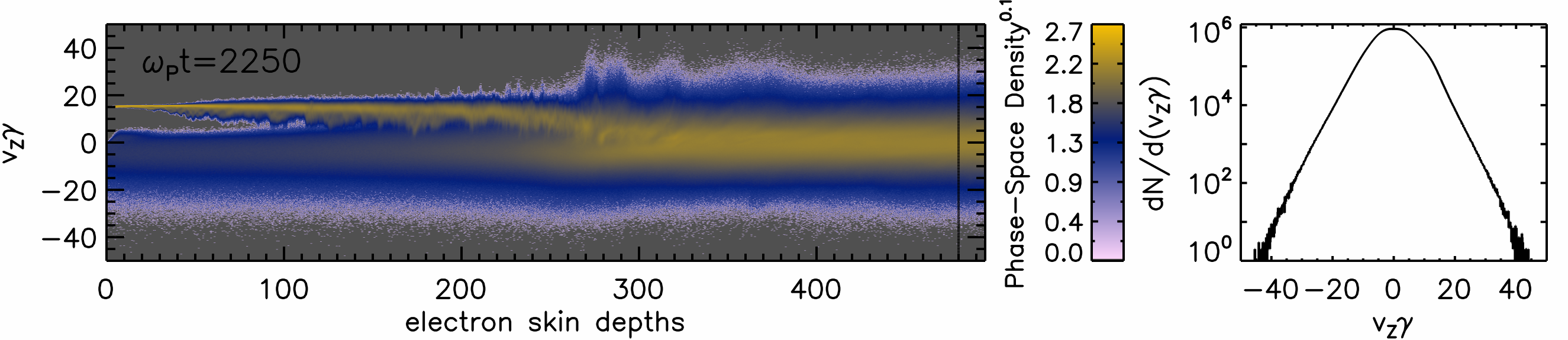}
\caption{Evolution of the ions in phase space. The PDF is sampled to the right of
the dotted line. Notice that it is only after $\omega_p t \simeq 2000$ that the downstream
part of the shock is completely thermalized and at rest. At earlier times the impact of
the wall is still significant, with the return current of reflected particles apparent in the
camel shaped pdf.}\label{fig:phasespace}
\end{figure}
reflecting cold streaming particles on a wall it takes some time until a proper, thermalized
downstream region is created. The electromagnetic fields at and near the shock interface
have to build up to a sufficient level to scatter the particles, and the shock interface has to be far
enough away from the wall at the upper boundary, so that upstream particles are scattered by the
fields and thermalize thoroughly before they possibly reach the wall. This convergence
can be monitored by looking at the $v_z\gamma$ momentum near the wall. A camel
shaped PDF signals that proper pressure support in the downstream region has still not been
established. We find (see figure \ref{fig:phasespace}) that the simulation has to run to
$\wpet\,=2000$ or $\omega_{pi}\,t\,=500$ before a proper equilibrium is established, even though
the proper jump conditions are already established at $\wpet\,=1000$ (see figure \ref{fig:dens}).
Without the moving frame approach it would have required a box with a least 28000 cells in the
streaming direction to properly establish and thermalize the shock.

\section{Conclusions}
In this Letter we have studied for the first time the long time behavior of a fully developed
3D collisionless ion-electron shock. We find that the development of the shock structure
is similar to that of an electron-ion shock in a 2D model, but there are both qualitative
and quantitative differences of importance when making quantitative predictions from
shock models for observations.
Apart from the dimensionality, which gives rise to different shock jump
conditions, the extra degree of freedom changes the shock in a number of ways: 1) The
current channels upstream of the shock become more stable.
2) Very strong parallel electric fields along the field lines are created at the shock interface
giving a new and effective avenue for particle acceleration, as compared to 2D models.
3) The level of magnetic field energy is lower near the shock interface
4) A power-law tail in the PDF downstream of the shock emerges at late times. In agreement with
theory the power-law index is shallower in the 3D shock ($\sim$2.2) compared to the 2D model ($\sim$2.5).

In \citet{Trier:2010} it was found that the radiation emitted from the Weibel instability in
2D and 3D is qualitatively different, and analogously we expect that the differences presented
here in the physical development of a 3D shock compared to a 2D shock will give rise to
significant differences in the emitted radiation.   This is a topic for future studies.

\acknowledgements TH acknowledges support from the Danish Natural Science Research Council.
Computer time was provided by the Danish Center for Scientific Computing (DCSC).
It is a pleasure to thank J.~T.~Frederiksen, M.~Medvedev, and \AA.~Nordlund for many
valuable discussions and comments.

\bibstyle{apj}
\bibliography{references}

\end{document}